\documentclass[a4paper,11pt]{article}

\usepackage{amssymb}
\usepackage{amsthm}
\usepackage{tikz}
\usepackage{tikz,pgfplots}
\usepackage{tkz-graph}
\usepackage{tkz-berge}
\usetikzlibrary{decorations.markings}
\usepackage{amsmath,amssymb}
\usepackage{todonotes}
\usepackage[T1]{fontenc}
\pgfplotsset{compat=1.18}
\usepackage{cite}
\usepackage{color}
\usepackage{hyperref}

\hypersetup{colorlinks=true}

\hypersetup{colorlinks=true, linkcolor=blue, citecolor=blue,urlcolor=blue}

\usepackage{graphicx}
\usepackage{placeins,caption}

\usepackage[top=3cm,bottom=3cm,left=2.8cm,right=2.8cm]{geometry}
	\newtheorem{theorem}{Theorem}[section]
	\newtheorem{Lemma}[theorem]{Lemma} 
	\newtheorem{Corollary}[theorem]{Corollary}  
	\newtheorem{proposition}[theorem]{Proposition}

	\theoremstyle{definition}

        \newtheorem{claim-proof}{Claim}
	
	\newtheorem{Remark}[theorem]{Remark} 

\begin{document}

\title{Roman domination in weighted graphs}
\author{Mart\'in Cera$^{a}$\thanks{Email: \texttt{mcera@us.es} ORCID: 0000-0002-3343-736X}
\and
Pedro Garc\'ia-V\'azquez$^{a}$\thanks{Email: \texttt{pgvazquez@us.es} ORCID:0000-0001-9208-9409}
\and
Juan Carlos Valenzuela-Tripodoro$^{b}$\thanks{Email: \texttt{jcarlos.valenzuela@uca.es} ORCID:0000-0002-6830-492X}
}

\maketitle

\begin{center}
$^a$ Departamento de Matem\'atica Aplicada I, Universidad de Sevilla, Spain \\
	\medskip
$^b$ Departamento de Matem\'aticas, Universidad de C\'adiz, Algeciras Campus, Spain
    \\
	\medskip
\end{center}

\begin{abstract}
A Roman dominating function for a (non-weighted) graph $G=(V,E)$, is a function $f:V\rightarrow \{0,1,2\}$ such that every vertex $u\in V$ with $f(u)=0$ has at least {one} 
neighbor $v\in V$ such that $f(v)=2$. The minimum weight $\sum_{v\in V}f(v)$ of a Roman {dominating function} $f$ on $G$ is called the Roman domination number of $G$ and is denoted by $\gamma_{R}(G)$. A graph 
{$G= (V,E)$} together with a positive real-valued weight-function $w:V\rightarrow \mathbf{R}^{>0}$ is called a {\it weighted graph} and is denoted by $(G;w)$. The minimum weight $\sum_{v\in V}f(v)w(v)$ of a Roman {dominating function} $f$ on $G$ is called the weighted Roman domination number of $G$ and is denoted by $\gamma_{wR}(G)$. The domination and Roman domination numbers of unweighted graphs have been extensively studied, particularly for their applications in bioinformatics and computational biology. However, graphs used to model biomolecular structures often require weights to be biologically meaningful. In this paper, we initiate the study of the weighted Roman domination number in weighted graphs. We first establish several bounds for this parameter and present various realizability results. Furthermore, we determine the exact values for several well-known graph families and demonstrate an equivalence between the weighted Roman domination number and the differential of a weighted graph.
\end{abstract}




\maketitle

\section{Introduction}
A (non-weighted) graph $G$ is a pair $(V,E)$, where $V$ is a set of vertices, and $E\subseteq V\times V$ is a set of edges, which are connections between those vertices, usually represented as pairs of vertices. A graph $G = (V,E)$ together with a positive real-valued weight-function $w:V\rightarrow \mathbf{R}^{>0}$ is called a {\it vertex-weighted graph} or simply {\it weighted graph}, and is denoted by $(G;w)$.

Graph Theory has established itself as a {versatile} 
and essential tool across multiple disciplines due to its unparalleled capacity to model any network in a structured and mathematical manner. This process of abstraction allows researchers to deepen their understanding of a system's behavior, evaluate its operational capacities, and develop robust predictive models. To achieve this, the practical challenge must first be formulated as a formal mathematical problem, centered on the analysis of specific graph parameters such as connectivity, centrality, or robustness.

In the majority of real-world scenarios, the components of a network and their interconnections are not uniform. They possess intrinsic properties that dictate the dynamics of the entire system. In Chemistry and Molecular Biology, the atoms within a molecule possess distinct masses and electronegativities \cite{N21, SNMT91} that influence molecular stability and reactivity. When modeling a protein as a graph, where nodes represent amino acids, the weight of an edge serves as a critical indicator of the physical distance or the interaction strength between them \cite{G11, AY12}. This spatial information is vital for understanding protein folding and enzymatic functions. In Communication Networks, unlike simple graphs where a connection is either "on" or "off", real digital networks depend on bandwidth \cite{CJWHFGW19}, latency \cite{ZRCZ21}, and throughput \cite{BCRSZZ12}. Weighting an edge allows engineers to model the cost of sending data through a specific route, which is the basis for modern internet routing protocols.

Despite the significant advances in Graph Theory and its widespread application in complex networks, much of the current theoretical body of work is based on the study of unweighted graphs. While these models are mathematically elegant, their scope is often limited when applied to the inherent heterogeneity of real-world networks. A model that treats all connections as equal can overlook critical bottlenecks or fail to identify the most efficient paths in a resource-constrained environment.

The transition from unweighted to weighted graph theory is not a simple task of direct translation. Some fundamental properties do not generalize easily. However, other properties show remarkable resilience and can be adapted through advanced mathematical frameworks. 

One of the most studied and applicable topics in the context of unweighted graphs is domination in graphs (see \cite{HHH23}). In its simplest form, a dominating set is a subset of nodes such that every node in the network is either in the set or adjacent to at least one member of the set. In a weighted interconnection network, the challenge evolves into finding a Minimum Weight Dominating Set. This has profound implications, for instance, for Ecology by determining the critical habitat patches that must be protected to ensure the survival of a species across a fragmented landscape \cite{BBE13}; and for Communication when we want to optimize the placement of sensors or servers to ensure total coverage of a network at the lowest possible infrastructure cost \cite{KB21}.

Roman domination is one of the most intriguing and widely studied variants within the field of domination in {Graph Theory}. Unlike classical domination, where a node simply covers its neighbors, Roman domination introduces the concept of strategic redundancy. This model does not merely seek coverage; it guarantees that the node providing protection retains its own integrity and functionality after deploying its resources to assist others. The concept is rooted in a military defense strategy proposed by the Emperor Constantine the Great, to protect the Roman Empire in the 4th century. Constantine decreed that no legion could leave its current location to assist a neighboring region under attack unless there was a second legion present to defend the original site. In mathematical terms, this problem was formalized in 1990 by Stewart \cite{S99} and the notion of Roman domination in graphs was introduced by Cockayne et al. \cite{CDHH04}. A Roman dominating function (RDF for short) on a non-weighted graph $G=(V,E)$ is defined as a function $f:V\rightarrow \{0,1,2\}$ such that every vertex $u\in V$ with $f(u)=0$ is adjacent at least to one vertex $v\in V$ for which $f(v)=2$. In other words, every RDF partitions the vertices into three sets $(V_0, V_1, V_2)$ based on the assigned label: $V_2$ is the subset of vertices with maximum resource capacity. $V_1$ is the subset of vertices with enough resources for their own defense, but which are incapable of assisting others. And $V_0$ is the subset of vertices without their own resources. To be considered protected or secured, they must be adjacent to at least one vertex in $V_2$. The aim is to find a RDF which ensures the entire graph is protected with the minimum cost. The significance of Roman domination lies in its utility for systems where resource mobility and service continuity are critical requirements. These are the cases of server architectures and data centers, emergency logistics and public safety projects and especially of systems biology and epidemic control, where it is used to identify control nodes within metabolic or protein interaction networks. If a drug must intervene in a biological pathway, Roman domination helps select intervention points that ensure the stability of the cellular system even under conditions of external stress or genetic mutation. 
{Several studies have focused on obtaining properties}
of this parameter and some of its variants (see for instance \cite{AAM24, BZ25, FKKS09, H02, H03, XYB09}).

In this research, we investigate the Roman domination number of a weighted graph. Our study begins by establishing a formal upper bound expressed in terms of the classical domination number. Furthermore, we derive and prove several general bounds that are demonstrated to be optimal. Finally, we determine the exact values of this parameter for various fundamental families of graphs, offering a precise reference for future applications in network topology and design.

\section{Definitions and terminology}
Throughout this article, only undirected simple graphs without loops or multiple edges are considered. Unless otherwise stated, we follow References \cite{ChJVZ24, GYZ13, HHH23} for terminology and deﬁnitions. 

A (non-weighted) graph $G$ is a pair $(V,E)$, where $V$ is a set of vertices, and $E\subseteq V\times V$ is a set of edges, which are connections between those vertices, usually represented as pairs of vertices. The neighborhood of a vertex $u\in V$ of $G$ is $N(u) = \{v\in V:(u,v)\in E\}$. The closed neighborhood of $u$ is $N[u]=N(u)\cup\{u\}$.

A graph 
{$G = (V,E)$}
together with a positive real-valued weight-function $w:V\rightarrow \mathbf{R}^{>0}$ is called a {\it vertex-weighted graph} or simply a {\it weighted graph}, and is denoted by $(G;w)$. Given a weighted graph $(G;w)$ and a subset $S\subseteq V$ the weight of $S$ is $w(S)=\sum_{v\in S}w(v)$. The {\it order of $G$} is $|V(G)|$ whereas the {\it weight of $G$} is $w(G)=w\left(V(G)\right)$. A weighted graph $(G;w)$ such that $w(G)=|V|$ is called a {\it normed weighted graph}. For every vertex $v \in V(G)$, the weighted degree of $v$ is defined as $d_w(v) = \frac{w(N(v))}{w(v)}$. Furthermore, let $\delta_w(G) = \min_{v \in V} \{d_w(v)\}$ and $\Delta_w(G) = \max_{v \in V} \{d_w(v)\}$ denote the {\it minimum} and {\it maximum weighted degree} of $G$, respectively. The subgraph of $G$ induced by $S$, denoted by $G[S]$, is the graph with vertex set $S$ whose edges consist of all edges in $G$ that connect two vertices in $S$. Analogously, the subgraph of $(G;w)$ induced by $S$, also denoted by $G[S]$, is the weighted graph $(G[S];w)$. In this case, the vertex and edge sets are defined identically to the unweighted case, and the weight function is restricted to $w:S\rightarrow \mathbf{R}^{>0}$. A subset of vertices $S \subseteq V$ is said to be independent if no two vertices in $S$ are adjacent. 

A weighted complete graph of order $n$, denoted by $(K_n;w)$, is a weighted graph in which every pair of distinct vertices is adjacent. It should be noted that, unlike unweighted graphs, a weighted complete graph is not unique. In fact, each weight function $w$ defines a distinct weighted complete graph.

Given a graph $G=(V,E)$ (resp. a weighted graph $(G;w)$), every vertex $v\in V$ dominates itself and each of its neighbors. We say that a subset $D\subseteq V$ is a dominating set (resp. weighted dominating set) of $G$ if every vertex of $V-D$ is neighbor of some vertex of $D$; that is, if $N[D]=V$. The minimum cardinality of a dominating set of $G$ is called the {\it domination number of $G$} and it is denoted by $\gamma(G)$. The minimum weight of a weighted dominating set of $G$ is called the {\it weighted domination number of $G$} and it is denoted by $\gamma_w(G)$.

In a graph $G : (V,E)$ (resp. weighted graph $(G;w)$), a {\it Roman dominating function}, for short RDF, (resp. {\it weighted Roman dominating function}, for short wRDF) 
is a mapping $f: V \longrightarrow \{0,1,2\}$ that satisfies the condition where each vertex 
$u$ with $f(u) = 0$ must be adjacent to at least one vertex $v$ such that $f(v) = 2$. The 
total weight of an RDF (resp. wRDF) is $f(V) = \sum_{u\in V} f(u)$ (resp. $f(V) = \sum_{u\in V} f(u)w(u)$). The smallest weight among all RDFs (resp. wRDFs) on $G$ (resp. $(G;w)$) is called the {\it Roman domination number} of $G$ (resp. {\it weighted Roman domination number} of $(G;w)$) and is denoted by $\gamma_R(G)$ (resp. $\gamma_{wR}(G)$).

\section{General bounds on weighted Roman domination number}
We first establish both upper and lower bounds for the weighted Roman domination number in terms of the weighted domination number.

\begin{theorem}\label{romcotgen}
    For every weighted graph $(G,w)$, it follows that $\gamma_{w}(G)\le\gamma_{wR}(G)\le 2\gamma_{w}(G)$.
\end{theorem}
\begin{proof}
{First,}
  let us show that $\gamma_{w}(G)\le\gamma_{wR}(G)$. Let $f$ be a $\gamma_{wR}$-function and consider the associated partition $(V_{0},V_{1},V_{2})$. Since every vertex of $V_{0}$ has at least one strong neighbor in $V_{2}$, we can ensure that $D=V_{1}\cup V_{2}$ is a weighted dominating set on $(G;w)$; 
    $$
        \begin{array}{rcl}
            \displaystyle \gamma_w(G) \le w(D) & = &  
            \displaystyle \sum_{u\in V_1}w(u)+\sum_{u\in V_2}w(u) 
                \le  \displaystyle \sum_{u\in V_1}w(u)+2\sum_{u\in V_2}w(u) \\[2ex] 
                & = & \displaystyle \sum_{i=0}^2 \sum_{u\in V_i} f(u) w(u)
                 =  f(V)=\gamma_{wR}(G).
        \end{array}$$
    
    {Second,}
    we prove that $\gamma_{wR}(G)\le 2\gamma_{w}(G)$. Let $D$ be a $\gamma_{w}(G)$-set of $(G;w)$ and consider the function $f:V\rightarrow\{0,1,2\}$ defined as $f(u)=2$ if $u\in D$, and $f(u)=0$ otherwise. It is evident that $f$ constitutes a wRDF on $(G;w)$. Consequently, we obtain: 
    $\displaystyle \gamma_{wR}(G)\le f(V)=\sum_{u\in D}2w(u)=2w(D)=2\gamma_w(G)$.
\end{proof}

Next, we examine the classes of weighted graphs where the equality $\gamma_{w} = \gamma_{wR}$ holds. Before doing so, we state the following necessary remark.

\begin{Remark}\label{Garifue}
Let $f = (V_0, V_1, V_2)$ be a $\gamma_{wR}$-function on $(G;w)$. If $f$ is chosen such that the cardinality of $V_1$ is minimized, then $V_1$ is an independent set in $(G;w)$. 
\end{Remark}
\begin{proof}
Let $f = (V_0, V_1, V_2)$ be a $\gamma_{wR}$-function that minimizes $|V_1|$. To prove the property by contradiction, suppose there exists an edge $(u,v)$ connecting two vertices in $V_1$. Without loss of generality, we may assume that $w(u) \le w(v)$. We then define a new function $g: V \to \{0, 1, 2\}$ as follows: $g(u) = 2$, $g(v) = 0$, and $g(z) = f(z)$ for all $z \in V \setminus \{u, v\}$. Since $u$ is assigned a value of 2, it is self-dominated and also dominates $v$. Consequently, $g$ is a weighted Roman dominating function with associated partition $(V_0\cup\{v\},V_1\setminus\{u,v\},V_2\cup \{u\})$. The total weight of $g$ satisfies $g(V) = f(V) + w(u) - w(v) \le f(V)$. However this fact contradicts the minimality of $|V_1|$ in $f$. Therefore, no such edge $(u,v)$ can exist, implying that $V_1$ is an independent set.
\end{proof}

We now characterize the classes of weighted graphs for which the lower bound of $\gamma_{wR}$ established in Theorem~\ref{romcotgen} is attained.

\begin{theorem}
Let $(G;w)$ be a weighted graph of order $n$. The equality $\gamma_w(G) = \gamma_{wR}(G)$ holds if and only if $G$ is the empty graph $\overline{K_n}$.
\end{theorem}
\begin{proof}
     If $(G, w)$ has no edges, then the only weighted dominating set is $D = V$, and consequently $w(V) = \gamma_w(G)$. Furthermore, the unique $\gamma_{wR}$-function $f$ on $(G, w)$ is defined by $f(v) = 1$ for all $v \in V$, with a weight of $\gamma_{wR}(G) = f(V) = w(V) = \gamma_w(G)$. Now, assume that $\gamma_w(G) = \gamma_{wR}(G)$. According to Remark~\ref{Garifue}, it suffices to show that the only $\gamma_{wR}$-function on $(G, w)$ is $f(v) = 1$ for all $v \in V$. Let $f$ be a $\gamma_{wR}$-function on $G$ and consider its associated partition $(V_0, V_1, V_2)$. We aim to prove that $V_1 = V$. Since every vertex $u$ with $f(u) = 0$ must be adjacent to at least one vertex $v$ with $f(v) = 2$, it follows that $V_1 \cup V_2$ is a weighted dominating set of $(G, w)$.
    Then 
    $$
    \begin{array}{rcl}
        \displaystyle \gamma_{wR}(G)=\gamma_w(G) & \le & w(V_1\cup V_2) 
                         =  \displaystyle \sum_{u\in V_1}w(u)+\sum_{u\in V_2} w(u)  \\[2ex] 
                        & \le & \displaystyle \sum_{u\in V_1}w(u)+2\sum_{u\in V_2}w(u)
                         =  \displaystyle \sum_{i=0}^2 \sum_{u\in V_i}f(u)w(u) 
                         =  f(V)= \gamma_{wR}(G),
    \end{array}
    $$ 
    and therefore 
    \begin{equation}\label{V2ais} 
        \displaystyle \sum_{u\in V_2} w(u)=0.
    \end{equation} 

This implies that $V_2 = \emptyset$ and, consequently, $V_0 = \emptyset$. Thus, $V = V_1$, which completes the proof.
\end{proof}

The following theorem upper bounds $\gamma_{wR}$ in terms of the graph weight and the maximum weighted degree $\Delta_w$. A similar bound for the weighted domination number was given in \cite{DRV04}.

\begin{theorem}\label{romcot}
Let $(G;w)$ be a non-trivial weighted graph. Then 
$$ \gamma_{wR}(G)\ge\left\lceil \frac{2w(G)}{\Delta_w+1}\right\rceil.$$ 
\end{theorem}
\begin{proof}
    Since $(G;w)$ is non-trivial, let $v^* \in V$ be a vertex such that $w(v^*) = \min\{w(v) : v \in V, N(v) \neq \emptyset\}$. It follows that $w(N(v^*)) \ge w(v^*)$, and thus $\Delta_w \ge 1$. Let $f: (G;w) \rightarrow \{0,1,2\}$ be a $\gamma_{wR}$-function with the associated partition $(V_0, V_1, V_2)$. For any $u \in V_2$, we observe that $\sum_{v \in N(u) \cap V_0} w(v) \le w(N(u)) \le \Delta_w w(u)$. Summing over all $u \in V_2$, we obtain:
    \begin{equation}\label{Delta_w}
        \sum_{u \in V_2}\left(\sum_{v \in N(u) \cap V_0} w(v)\right) \le \Delta_w \sum_{u \in V_2} w(u).
    \end{equation}
    Then,  it follows from (\ref{Delta_w}) that 
    $$
    \begin{array}{rcl} 
(\Delta_w+1)\gamma_{wR}(G) & = & \displaystyle (\Delta_w+1)\sum_{u\in V_1}w(u)+(\Delta_w+1)\sum_{u\in V_2}2w(u)\\[1ex]
& \ge & \displaystyle (\Delta_w+1)\sum_{u\in V_1}w(u)+2\sum_{u\in V_2}w(u)+2\sum_{u \in V_2}
\left( \sum_{v \in N(u) \cap V_0} w(v)\right) \\[1ex] 
& \ge & \displaystyle 2\sum_{u\in V_1}w(u)+2\sum_{u\in V_2}w(u)+2\sum_{v \in V_0} w(v)\\[1ex] & = & 2w(G).
\end{array}
$$ Hence, $\displaystyle\gamma_{wR}(G)\ge\left\lceil\frac{2w(G)}{\Delta_w+1}\right\rceil$, which finishes the proof.
\end{proof}

For unweighted graphs, it was proven in \cite{CDHH04} that $\gamma_R \ge \frac{2n}{\Delta+1}$, where $n$ denotes the order and $\Delta$ the maximum degree of the graph. The bound established in Theorem \ref{romcot} serves as a weighted generalization of this result. Indeed, if a weighted graph $(G;w)$ of order $n$ is normed, then $w(G)=n$. Consequently, Theorem \ref{romcot} implies that $\gamma_{wR}(G) \ge \frac{2n}{\Delta_w+1}$.

\begin{figure}
    \centering
    \includegraphics[width=0.5\linewidth]{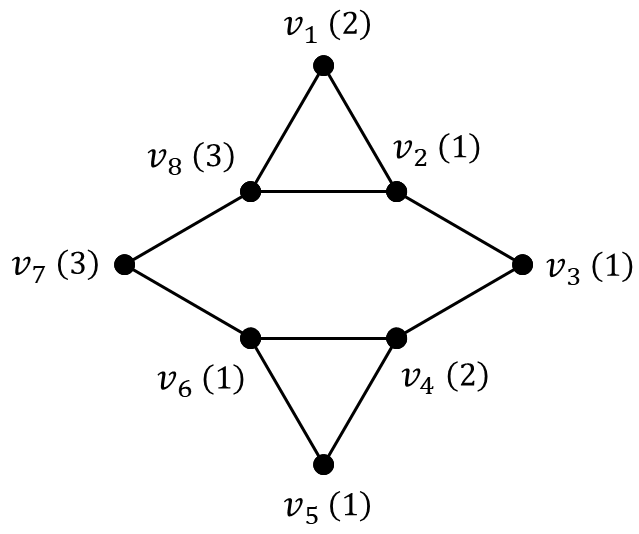}
    \caption{An example for which the upper bound of Theorem \ref{romcot} is reached.}
    \label{fig:uppstar}
\end{figure}

The upper bound of Theorem \ref{romcot} is tight, as we can see in the weighted graph $(G;w)$ depicted in Figure~\ref{fig:uppstar}. Indeed, $\gamma_{wR}(G)=4$, and the function $f:(G;w)\rightarrow\{0,1,2\}$, defined by $f(v_2)=f(v_6)=2$ and $f(v_i)=0$ for $i \notin \{1, 6\}$, is a $\gamma_{wR}$-function. Furthermore, based on the vertex weights (indicated in parentheses), we have $\Delta_w=6$ and $w(G)=14$. Consequently, the following equality holds: $\gamma_{wR}(G)=\frac{2w(G)}{\Delta_w+1}$. There are also some families of weighted graphs for which the upper bound of Theorem \ref{romcot} is reached, as we will see in the next section.

The following proposition establishes several properties of the partition associated with a $\gamma_{wR}$-function.

\begin{proposition}\label{propiedades}
     Let $(G;w)$ be a weighted graph and let $f$ be a $\gamma_{wR}$-function with associated partition $(V_0, V_1, V_2)$. Then, the following assertions hold:
     \begin{itemize}
         \item[(a)] If $u, v \in V_1$ are adjacent, then $w(u) = w(v)$.
         \item[(b)] No edge of $G$ joins a vertex in $V_1$ to a vertex in $V_2$.
         \item[(c)] $V_2$ is a $\gamma_w$-set of the induced subgraph $G[V_0 \cup V_2]$.
     \end{itemize}
\end{proposition}
\begin{proof}
    (a) Without loss of generality, assume that $w(u) \le w(v)$. Consider the function $g: V \rightarrow \{0,1,2\}$ defined by $g(u)=2$, $g(v)=0$, and $g(z)=f(z)$ for all $z \in V \setminus \{u, v\}$. It is clear that $g$ is a wRDF on $(G;w)$. Its weight is given by $$g(V) = f(V) - (w(u) + w(v)) + 2w(u) = f(V) + w(u) - w(v) \le f(V).$$ Since $f$ is a $\gamma_{wR}$-function, it must follow that $w(u) = w(v)$ to avoid a contradiction with the optimality of $f$. 

    (b) Suppose on the contrary there exists an edge joining $u \in V_1$ and $v \in V_2$. By setting $f(u) = 0$ and leaving all other values unchanged, we obtain a new wRDF on $(G;w)$ with a strictly smaller weight than $f$, which contradicts the optimality of $f$.

    (c) Since each vertex $u \in V_0$ has at least one neighbor in $V_2$, it follows that $V_2$ is a weighted dominating set of the induced subgraph $G[V_0 \cup V_2]$. Suppose $V_2$ is not a $\gamma_w$-set of $G[V_0 \cup V_2]$. Then, there exists a weighted dominating set $D \subseteq V_0 \cup V_2$ such that $w(D) < w(V_2)$. In this case, we define a new function $g: V \rightarrow \{0,1,2\}$ as follows: $g(v)=2$ if $v \in D$, $g(v)=1$ if $v \in V_1$, and $g(v)=0$ otherwise. This function $g$ is a wRDF on $(G;w)$ with an associated partition $((V_0 \cup V_2) \setminus D, V_1, D)$. However, its weight satisfies: $$g(V) = 2w(D) + w(V_1) < 2w(V_2) + w(V_1) = f(V),$$ which contradicts the optimality of $f$ as a $\gamma_{wR}$-function.  
\end{proof}

Next we will provide a condition that is both necessary and sufficient for a non-trivial weighted graph to have a $\gamma_{wR}$ equal to its weight. Before, we need to prove that every two neighbors must have the same weight. 

\begin{Lemma}\label{isolaedge}
    Let $(G;w)$ be a non-trivial weighted graph. Assume that $\gamma_{wR}(G) = w(G)$ and let $f$ be a $\gamma_{wR}$-function on $(G;w)$ with the associated partition $(V_0, V_1, V_2)$. Then, the following assertions hold:
    \begin{itemize}
        \item[(i)] For every $u \in V_2$, $w(u) = \sum_{v \in N(u) \cap V_0} w(v)$.
        \item[(ii)] $V_2$ is an independent set.
        \item[(iii)] If $u \in V_2$ and $v \in V_0$ are adjacent, then the edge $(u,v)$ is a connected component of $(G;w)$.
    \end{itemize} 
\end{Lemma}
\begin{proof}
$(i)$ Suppose, for the sake of contradiction, that there exists $u \in V_2$ such that $w(u) \neq \sum_{v \in N(u) \cap V_0} w(v)$.

Case 1: Assume $w(u) < \sum_{v \in N(u) \cap V_0} w(v)$. Define a function $g: V \rightarrow \{0,1,2\}$ such that $g(u)=2$, $g(v)=0$ for all $v \in N(u) \cap V_0$, and $g(v)=1$ otherwise. It is clear that $g$ is a wRDF on $(G;w)$. However, its weight satisfies:$$\begin{aligned} g(V) &= \sum_{v \in V \setminus (N(u) \cap V_0 \cup \{u\})} w(v) + 2w(u)  \\ &< \sum_{v \in V \setminus (N(u) \cap V_0 \cup \{u\})} w(v) + w(u)+\sum_{v \in N(u) \cap V_0} w(v) = w(G) \end{aligned}$$ which contradicts the assumption that $\gamma_{wR}(G) = w(G)$.

Case 2: Assume $w(u) > \sum_{v \in N(u) \cap V_0} w(v)$. In this case, we have: $$\begin{aligned} w(G) = f(V) &= \sum_{v \in V \setminus (N(u) \cap V_0 \cup \{u\})} f(v)w(v) + 2w(u) \\ &> \sum_{v \in V \setminus (N(u) \cap V_0 \cup \{u\})} f(v)w(v) + w(u) + \sum_{v \in N(u) \cap V_0} w(v) = g(V). \end{aligned}$$ This implies $g(V) < f(V)$, contradicting the fact that $f$ is a $\gamma_{wR}$-function.

Consequently, $w(u) = \sum_{v \in N(u) \cap V_0} w(v)$ must hold for every $u \in V_2$.

$(ii)$ If $u, v \in V_2$ are adjacent, we can reassign $g(v)=0$ and $g(z)=1$ for all $z \in N(v) \cap V_0$. Since $u$ dominates $v$, $g$ remains a wRDF. By item $(i)$, we know that $\sum_{z \in N(v) \cap V_0} w(z)=w(v)$. Thus, $g(V)=f(V) - 2w(v) + \sum_{z \in N(v) \cap V_0} w(z) = f(V) - w(v) < w(G)$, a contradiction. Hence, $V_2$ must be an independent set.

    $(iii)$ Let $u \in V_2$ and $v \in V_0$ be adjacent vertices. By Proposition \ref{propiedades} and item $(ii)$, it follows that $N(u) \subseteq V_0$. From item $(i)$, we have the identity:
    \begin{equation}\label{u^*}
    w(u) = \sum_{z \in N(u)} w(z),
    \end{equation} 
    which particularly implies:
    \begin{equation}\label{u>v}
    w(u) \ge w(v).
    \end{equation}
    Let $g: V \rightarrow \{0,1,2\}$ be the function defined by $g(u)=0$, $g(v)=2$, $g(z)=1$ for all $z \in N(u) \setminus \{v\}$, and $g(z)=f(z)$ otherwise. It is straightforward to verify that $g$ is a wRDF on $(G;w)$. Based on equations \eqref{u^*} and \eqref{u>v}, we obtain:$$\begin{aligned}
\gamma_{wR}(G) = f(V) &= \sum_{z \in V \setminus N[u]} f(z)w(z) + 2w(u) \\
&= \sum_{z \in V \setminus N[u]} g(z)w(z) + w(u) + \sum_{z \in N(u)} w(z) \\
&= \sum_{z \in V \setminus N[u]} g(z)w(z) + (w(u) - w(v)) + \sum_{z \in N(u)} g(z)w(z) \\
&= g(V) + w(u) - w(v) \\
&\ge g(V) \ge \gamma_{wR}(G).
\end{aligned}$$ This chain of inequalities implies that $w(u) = w(v)$, which, according to \eqref{u^*}, is only possible if $N(u) = \{v\}$. Furthermore, as $g(V) = f(V)$, the function $g$ is also a $\gamma_{wR}$-function with $g(v)=2$ and $g(u)=0$. Applying the same reasoning to $v$ leads to $N(v) = \{u\}$. Consequently, every edge connecting $V_2$ and $V_0$ is a connected component.
\end{proof}

\begin{theorem}\label{thp}
    Let $(G;w)$ be a non-trivial weighted graph. Then $\gamma_{wR}(G) \le w(G)$. Equality holds if and only if every connected component of $G$ is either an isolated vertex or a single edge $(x_i,y_i)$ such that $w(x_i) = w(y_i)$. 
\end{theorem}
\begin{proof}
    Clearly, $\gamma_{wR}(G) \le w(G)$, as the function $f: V \rightarrow \{0,1,2\}$ defined by $f(v)=1$ for all $v \in V$ is a wRDF with weight $\omega(f) = \sum_{v \in V} w(v) = w(G)$. Suppose that every connected component of $G$ is either an isolated vertex or a single edge. Let the set of edges be $E = \{\{x_1, y_1\}, \dots, \{x_r, y_r\}\}$ and let $Z$ be the set of isolated vertices, such that $V = X \cup Y \cup Z$, where $X=\{x_1, \dots, x_r\}$ and $Y=\{y_1, \dots, y_r\}$. Given that $w(x_i) = w(y_i)$ for all $i$, it follows directly that $\gamma_{wR}(G) = w(G)$.
    
    Conversely, assume $\gamma_{wR}(G) = w(G)$, and let $f$ be a $\gamma_{wR}$-function with associated partition $(V_0, V_1, V_2)$. By the definition of weighted Roman domination, no vertex in $V_1$ can be adjacent to a vertex in $V_2$; otherwise, reassigning $f(v)=0$ for the vertex $v \in V_1$ would yield a wRDF of smaller weight, a contradiction. Furthermore, by Lemma \ref{isolaedge}, we deduce that each vertex has at most one neighbor, and that any two adjacent vertices must have the same weight. This completes the proof.
\end{proof}

As a consequence of Theorem \ref{thp}, we establish a Nordhaus–Gaddum-type inequality for the sum of the weighted Roman domination numbers of a weighted graph and its complement.

\begin{proposition}\label{nord}
Let $(G;w)$ be a weighted graph, with $n\ge 3$ vertices, such that both $(G;w)$ and $(\overline{G};w)$ are non-trivial. Then,$$\displaystyle 4\min_{v\in V}\{w(v)\} \le \gamma_{wR}(G) + \gamma_{wR}(\overline{G}) < w(G) + w(\overline{G}).$$
\end{proposition}

\begin{proof}
Let $f$ be a $\gamma_{wR}$-function on $(G;w)$ that minimizes the cardinality of $V_1$. Since $(G;w)$ is a non-trivial graph, it contains at least one edge. This implies that $V_2 \neq \emptyset$, and consequently, $\gamma_{wR}(G) = \omega(f) \ge 2 \min_{v \in V}\{w(v)\}$. Applying the same reasoning to the complement, we obtain $\gamma_{wR}(\overline{G}) \ge 2 \min_{v \in V}\{w(v)\}$. Summing these two lower bounds yields the first inequality. Regarding the upper bound, Theorem \ref{thp} establishes that $\gamma_{wR}(G) \le w(G)$ and $\gamma_{wR}(\overline{G}) \le w(\overline{G})$. According to the same theorem, equality $\gamma_{wR}(G) = w(G)$ holds only if every connected component of $G$ is either an isolated vertex or an isolated edge. Crucially, a graph and its complement cannot simultaneously possess this structure. Therefore, at least one of the inequalities must be strict: either $\gamma_{wR}(G) < w(G)$ or $\gamma_{wR}(\overline{G}) < w(\overline{G})$. It follows that:$$\gamma_{wR}(G) + \gamma_{wR}(\overline{G}) < w(G) + w(\overline{G}).$$
\end{proof}

Since $w(G) = w(\overline{G})$, the upper bound in Proposition \ref{nord} can be equivalently expressed as $\gamma_{wR}(G) + \gamma_{wR}(\overline{G}) < 2w(G)$.

\section{The weighted Roman domination number of some families of graphs}
In this section, we determine the weighted Roman domination number for complete graphs and complete bipartite graphs. Furthermore, we provide approximations for this parameter in weighted cycles.

The following result is a direct consequence of Theorem \ref{romcot}. 

\begin{Corollary}\label{complete}
    For a complete weighted graph $(K_n;w)$ of order $n\ge 2$, 
    $$\gamma_{wR}(K_n)=2\min\{w(v):v\in V\}.$$
\end{Corollary}
\begin{proof}
  Let $u^* \in V$ be a vertex such that $w(u^*) = \min \{w(v) : v \in V\}$. We shall prove that $\gamma_{wR}(K_n) = 2w(u^*)$. First, consider the function $f: V \rightarrow \{0,1,2\}$ defined by $f(u^*) = 2$ and $f(v) = 0$ for all $v \in V \setminus \{u^*\}$. It is clear that $f$ is a wRDF, as $u^*$ is adjacent to all other vertices in $K_n$. The weight of this function is $f(V) = 2w(u^*)$, which implies that $\gamma_{wR}(K_n) \le 2w(u^*)$. To establish the lower bound, let $\Delta_w$ be the weighted maximum degree. For a complete graph, $\Delta_w = \frac{w(K_n) - w(u^*)}{w(u^*)}$. By applying Theorem \ref{romcot}, we obtain: $$\gamma_{wR}(K_n) \ge \frac{2w(K_n)}{\Delta_w + 1} = \frac{2w(K_n)}{\frac{w(K_n)}{w(u^*)}} = 2w(u^*).$$ Combining both inequalities, we conclude that $\gamma_{wR}(K_n) = 2w(u^*)$.
\end{proof}

\begin{figure}[ht]
    \centering
    \includegraphics[width=0.9\linewidth]{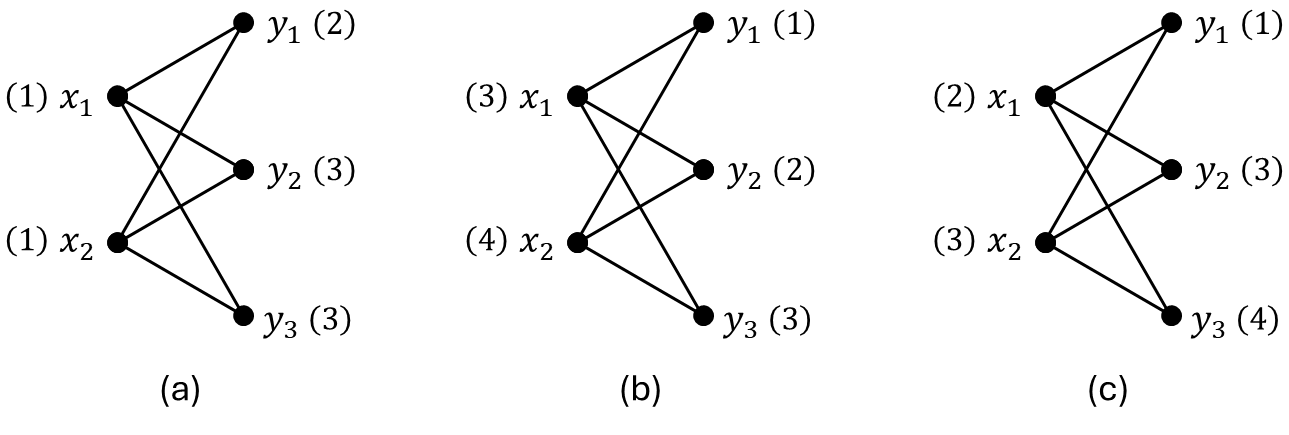}
    \caption{Three different strategies depending on the distribution of vertex weights.}
    \label{fig:bipart}
\end{figure}

In general, obtaining closed-form expressions for the exact value of the $\gamma_{wR}$ parameter is a challenging task, even when restricted to well-known families of weighted graphs. In the classical unweighted case, an optimal strategy for Roman domination in complete bipartite graphs $K_{2,r}$ is to assign labels from $\{1, 2\}$ to the vertices in the partition of cardinality two. However, in the weighted case, this strategy may not be optimal, as the vertex weights can favor a different distribution of values (see Figure \ref{fig:bipart}). Indeed, the examples in Figure \ref{fig:bipart} illustrate how the optimal weighted Roman dominating function (wRDF) adapts to different weight distributions. In graph \textbf{(a)}, the optimal wRDF assigns $f(x_1)=2$, $f(x_2)=1$, and $f(y_j)=0$ for $j=1,2,3$. In graph \textbf{(b)}, it is optimal to set $f(x_i)=0$ for $i=1,2$, while assigning $f(y_1)=2$ and $f(y_j)=1$ for $j=2,3$. In graph \textbf{(c)}, the minimum weight is achieved by setting $f(x_1)=f(y_1)=2$ and $f(x_2)=f(y_j)=0$ for $j=2,3$.

The following proposition provides the exact value of the weighted Roman domination number for a complete bipartite weighted graph. This value is expressed in terms of the total weights of both vertex classes and the minimum vertex weight within each class. 

\begin{proposition}
    Let $(K_{s,t}, w)$ be a complete bipartite weighted graph with vertex classes $X = \{x_1, \dots, x_s\}$ and $Y = \{y_1, \dots, y_t\}$, where $s \le t$. Assume the vertices are ordered by non-decreasing weight, such that $w(x_i) \le w(x_{i+1})$ and $w(y_j) \le w(y_{j+1})$. Then:
    \begin{itemize}
    \item[(i)] If $s=1$, then $\gamma_{wR}(K_{1,t}) = \min\{2w(x_1), w(y_1) + w(Y)\}$.
    \item[(ii)] If $s \ge 2$, then $\gamma_{wR}(K_{s,t}) = \min \{ w(x_1) + w(X), w(y_1) + w(Y), 2(w(x_1) + w(y_1)) \}.$
    \end{itemize}
\end{proposition}

\begin{proof}    
    Let $f = (V_0, V_1, V_2)$ be a $\gamma_{wR}$-function on $K_{s,t}$ that minimizes $|V_1|$. Since $w(x_i) \le w(x_{i+1})$ and $w(y_j) \le w(y_{j+1})$, we may assume without loss of generality that the labels are assigned in non-increasing order; that is, $f(x_i) \ge f(x_{i+1})$ and $f(y_j) \ge f(y_{j+1})$.
    
    By Remark \ref{Garifue}, the set $V_1$ must be entirely contained within either $X$ or $Y$. Additionally, Proposition \ref{propiedades} establishes that vertices in $V_1$ and $V_2$ cannot be adjacent. Given that $f$ is chosen to minimize $|V_1|$, these conditions imply that $f(x_1) \neq 1$ and $f(y_1) \neq 1$. Indeed, if a vertex were assigned label 1, any vertex with label 2 would be forced into the same partition class to avoid adjacency; however, such a configuration would contradict Remark \ref{Garifue} or the minimality of $|V_1|$.
    
    $(i)$ Case $s=1$: If $f(x_1) = 0$, then $x_1$ must be dominated by $Y$, which requires $f(y_1) = 2$ and $f(y_j) = 1$ for all $j=2, \dots, t$. On the other hand, if $f(x_1) = 2$, then all vertices in $Y$ are already dominated, so $f(y_j) = 0$ for all $j$. This leads to:$$ \gamma_{wR}(K_{1,t}) = \min\{ 2w(x_1), w(y_1) + w(Y) \}. $$
    
    $(ii)$ Case $s \ge 2$: We consider the possible assignments for $f(x_1)$ and $f(y_1)$. If $f(x_1) = 0$, then $x_1$ must be dominated by $Y$. The optimal choice is $f(y_1) = 2$. To ensure all $x_i$ are dominated when $f(x_i) = 0$, we must have $f(y_j) = 1$ for $j=2, \dots, t$. If $f(x_1) = 2$, all vertices in $Y$ are dominated. For the remaining vertices in $X$ ($i \ge 2$), there are two subcases: 
    
    \begin{itemize}
        \item They are assigned label 1, which implies $f(y_1) = 0$ (to maintain the non-adjacency between $V_1$ and $V_2$).
        \item They are assigned label 0, which requires $f(y_1) = 2$ to ensure their domination.
    \end{itemize}
    
    It follows that: $$ \gamma_{wR}(K_{s,t}) = \min \{ w(x_1) + w(X), w(y_1) + w(Y), 2(w(x_1) + w(y_1)) \}, $$ and the proof is complete.
\end{proof}

The Roman domination number of an unweighted cycle $C_n$ of order $n$ is well known and straightforward to determine; specifically, $\gamma_{R}(C_n) = \left\lceil \frac{2n}{3} \right\rceil$. However, when weights are assigned to the vertices, the problem becomes considerably more complex. In the following theorem, we provide an upper bound for the weighted Roman domination number in terms of the total weight of the cycle. Furthermore, for cases where $n \not\equiv 0 \pmod{3}$, we will prove that this bound is attained if and only if the weights are uniformly distributed among all vertices.

\begin{theorem}\label{cyclecota}
    Let $(C_n, w)$ be a weighted cycle of order $n \ge 3$, and let $k = \lfloor n/3 \rfloor$. Then:$$\gamma_{wR}(C_n) \le \left( 1 - \frac{k}{n} \right) w(C_n).$$
\end{theorem}
\begin{proof}
   Let the weighted cycle be denoted by $C_n: u_1, u_2, \dots, u_n, u_1$. We analyze the problem by considering three cases based on the remainder of $n$ modulo 3.
   
   {\it Case 1:} $n = 3k$. For each $m \in \{1, \dots, n\}$, define the function $f_m: V \rightarrow \{0, 1, 2\}$ as:$$ f_{m}(u_{(m+3i) \pmod n}) = 2 \quad \text{for } i = 0, \dots, k-1, $$and $f_m(v) = 0$ otherwise. Each $f_m$ is a weighted Roman dominating function (wRDF) on $C_n$. Summing the weights of all such functions, we obtain:
   \begin{equation}\label{cyine0}
   n \cdot \gamma_{wR}(C_n) \le \sum_{m=1}^n f_m(V) = 2k \sum_{j=1}^{n} w(u_{j}) = 2k w(C_n).
   \end{equation}
   It follows that $\gamma_{wR}(C_n) \le \frac{2k}{n} w(C_n) = \left( 1 - \frac{k}{n} \right) w(C_n)$.
   
   {\it Case 2:} $n = 3k + 1$. For each $m \in \{1, \dots, n\}$, define $f_m: V \rightarrow \{0, 1, 2\}$ as:$$ f_m(u_{m}) = 1, \quad f_{m}(u_{(m+2+3i) \pmod n}) = 2 \quad \text{for } i = 0, \dots, k-1, $$and $f_m(v) = 0$ otherwise. Since each $f_m$ is a wRDF on $C_n$, we have:
   \begin{equation}\label{cyine1}
   n \cdot \gamma_{wR}(C_n) \le \sum_{m=1}^n f_m(V) = (2k+1) \sum_{j=1}^{n} w(u_{j}) = (2k+1) w(C_n).
   \end{equation}
   Therefore, $\gamma_{wR}(C_n) \le \frac{2k+1}{n} w(C_n) = \left( 1 - \frac{k}{n} \right) w(C_n)$.
   
   {\it Case 3:} $n = 3k + 2$.For each $m \in \{1, \dots, n\}$, define $f_m: V \rightarrow \{0, 1, 2\}$ as:$$ f_{m}(u_{(m+3i) \pmod n}) = 2 \quad \text{for } i = 0, \dots, k, $$and $f_m(v) = 0$ otherwise. Clearly, $\{f_m : m=1, \dots, n\}$ is a set of wRDFs on $C_n$, which implies:
   \begin{equation}\label{cyine2}
   n \cdot \gamma_{wR}(C_n) \le \sum_{m=1}^n f_m(V) = 2(k+1) \sum_{j=1}^{n} w(u_{j}) = (2k+2) w(C_n).
   \end{equation}
   Thus, $\gamma_{wR}(C_n) \le \frac{2k+2}{n} w(C_n) = \left( 1 - \frac{k}{n} \right) w(C_n)$.
\end{proof}

\begin{theorem}\label{cycle:n_not_3k}
    Let $n, k \ge 1$ be integers such that $n = 3k + \ell$ for $\ell \in \{1, 2\}$. Let $(C_n, w)$ be a weighted cycle of order $n$. Then,$$ \gamma_{wR}(C_n) = \left( 1 - \frac{k}{n} \right) w(C_n) $$if and only if the weights of all vertices are equal (i.e., $w$ is a constant weight function).
\end{theorem}
\begin{proof}
   Let the weighted cycle be denoted by $C_n: u_1, u_2, \dots, u_n, u_1$. If all vertex weights are equal, say $w(u_i) = p$ for some constant $p > 0$, then $(C_n, w \equiv p)$ can be identified with an unweighted cycle $C'_n$ scaled by $p$. In this case, $\gamma_{wR}(C_n) = \gamma_R(C'_n)p = \lceil \frac{2n}{3} \rceil p = (2k+\ell)p$.
   
   Now, let us show the converse. Assume that $\gamma_{wR}(C_n) = (1 - \frac{k}{n})w(C_n)$. Under this assumption, all inequalities in Theorem~\ref{cyclecota} become equalities—specifically, those in (\ref{cyine1}) if $n=3k+1$, and those in (\ref{cyine2}) if $n=3k+2$. Consequently, every function in the set $\{f_m : m = 1, \dots, n\}$ is a $\gamma_{wR}(C_n)$-function. The set of $n$ identities $\{f_m(V) = \gamma_{wR}(C_n)\}_{m=1}^n$ can be expressed as a linear system of the form $\mathbf{A}\mathbf{w} = \mathbf{b}$, where:$$\mathbf{w} = \begin{pmatrix} w(u_1) \\ w(u_2) \\ \vdots \\ w(u_n) \end{pmatrix}, \quad \mathbf{b} = \begin{pmatrix} \gamma_{wR}(C_n) \\ \gamma_{wR}(C_n) \\ \vdots \\ \gamma_{wR}(C_n) \end{pmatrix},$$and $\mathbf{A}$ is a circulant matrix whose rows correspond to the labels assigned by each $f_m$. Specifically, the entries of each row are determined by the labels $\{0, 1, 2\}$ defined in the construction of the functions $f_m$. This linear system has a unique solution by the Rouché-Frobenius Theorem, provided that $\text{rank}(\mathbf{A}) = n$ (i.e., $\mathbf{A}$ is non-singular). We observe that the sum of the entries in each row of $\mathbf{A}$ is constant and equal to $2k+1$ if $n=3k+1$, and $2k+2$ if $n=3k+2$. Since the vector $\mathbf{b}$ is also constant, it follows that $w(u_i) = \frac{w(C_n)}{n}$ for all $i=1, \dots, n$ is a solution to the system. Given that the matrix is non-singular, this is the unique solution, which implies that all vertex weights are equal, thus completing the proof.
\end{proof}

\begin{figure}[ht]
    \centering
    \includegraphics[width=0.9\linewidth]{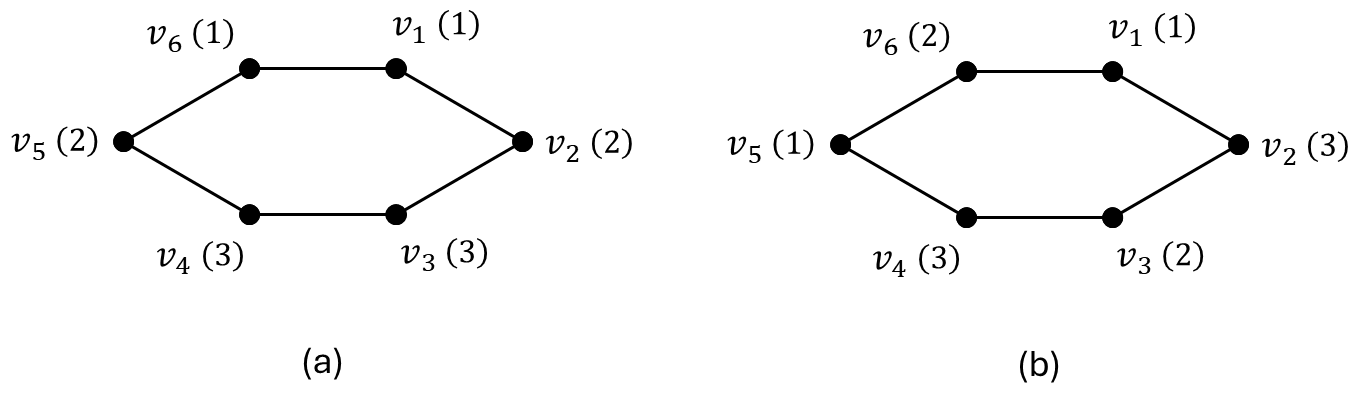}
    \caption{Two examples of weighted hexagons where the equality in Theorem \ref{cyclecota} holds.}
    \label{fig:hex}
\end{figure}

The reasoning used in the proof of Theorem \ref{cycle:n_not_3k} can be extended to the case $n=3k$ to demonstrate that $\gamma_{wR}(C_n) = \frac{2}{3}w(C_n)$ does not necessarily imply that all vertex weights are equal. This occurs because the circulant matrix associated with the linear system $\{f_m(V) = \gamma_{wR}(C_n)\}_{m=1}^n$, derived from equation (\ref{cyine0}) in Theorem \ref{cyclecota}, has a rank strictly less than $n$. Specifically, for $n=3k$, the rank of this matrix is 3 (for $n>3$). Consequently, when $n=3k$, there are infinitely many weighted graphs for which the bound in Theorem \ref{cyclecota} is attained. Figure \ref{fig:hex} illustrates this by showing two weighted hexagons ($(C_6;w)$) with the same total weight but different vertex weight distributions, both of which satisfy the equality in Theorem~\ref{cyclecota}. 

\section{The weighted Roman domination number and the differential of a weighted graph}
The concept of the differential of an unweighted graph was first introduced in \cite{MHHHS06}, with subsequent applications in the study of information diffusion within social networks. For an unweighted graph $G=(V,E)$ and a subset $S\subseteq V$, the {\it differential of $S$} is defined as $\partial (S) = |B(S)|-|S|$, where $B(S)$ denotes the set of vertices in $V\setminus S$ adjacent to at least one vertex in $S$. The {\it differential of the graph $G$} is then given by: $$\partial(G)=\max\{\partial(S):S\subseteq V\}$$ A subset $S\subseteq V$ is called a $\partial$-set (or {\it differential set}) if it satisfies $\partial (S) = \partial(G)$. 

This notion naturally extends to weighted graphs. For a weighted graph $(G, w)$ and a subset $S\subseteq V$, the differential of $S$ is defined as:$$\partial (S)=w(B(S))-w(S)$$where $w(X) = \sum_{v \in X} w(v)$. Similarly, the {\it differential of $(G, w)$} is defined as $\partial(G)=\max\{\partial(S):S\subseteq V\}$.

The following theorem establishes a relationship between the weighted Roman domination number and the differential of a weighted graph.

\begin{theorem}\label{dif}
    Let $(G;W))$ be a weighted graph. Then $\gamma_{wR}(G)=w(G)-\partial(G)$.
\end{theorem}
\begin{proof}
   Let $f$ be a $\gamma_{wR}(G)$-function with its associated partition $(V_0, V_1, V_2)$. By setting $S = V_2$, we observe that $B(S) = V_0$ and $V_1 = V \setminus (S \cup B(S))$. Consequently, we have:$$\begin{aligned}
\gamma_{wR}(G) &= 2w(V_2) + w(V_1) \\
&= w(V_2) + w(V_0) - \partial(V_2) + w(V_1) \\
&= w(G) - \partial(V_2) \\
&\ge w(G) - \max\{\partial(S) : S \subseteq V\} \\
&= w(G) - \partial(G).
\end{aligned}$$ To prove the reverse inequality, let $S^* \subseteq V$ be a subset such that $\partial(G) = \partial(S^*)$. We define a function $g: V \to \{0, 1, 2\}$ as follows: $g(v) = 2$ if $v \in S^*$, $g(v) = 0$ if $v \in B(S^*)$, $g(v) = 1$ otherwise. Since $g$ is a wRDF, it follows that: $$\begin{aligned}
\gamma_{wR}(G) \le g(V) &= 2w(S^*) + w(V \setminus (S^* \cup B(S^*))) \\
&= w(S^*) + w(B(S^*)) - \partial(S^*) + w(V \setminus (S^* \cup B(S^*))) \\
&= w(G) - \partial(S^*) \\
&= w(G) - \partial(G).
\end{aligned}$$ Thus, we conclude that $\gamma_{wR}(G) = w(G) - \partial(G)$.
\end{proof}

\section{Discussion}
While research on unweighted graphs is extensive, its application to real-world networks is often limited. Real-world structures typically consist of points of interest with varying degrees of relevance, which are more accurately modeled by graphs with vertex weights. In this paper, we have initiated the study of weighted Roman domination in weighted graphs. Specifically, we have established several bounds for the weighted Roman domination number in terms of graph invariants such as total weight, maximum weighted degree, and the weighted domination number.

Furthermore, we have determined the exact values of $\gamma_{wR}$ for specific families of graphs, including complete weighted graphs, complete bipartite weighted graphs, and weighted cycles. Finally, we have established a fundamental relationship between the weighted Roman domination number and the differential of a weighted graph.

As a continuation of this research, we suggest the following lines of investigation:
\begin{itemize}
\item Establish new lower bounds for $\gamma_{wR}$ in terms of the minimum weighted degree $\delta_w$.
\item Building upon Theorem \ref{dif}, seek further refinements of the general bounds presented in this work.
\end{itemize}

\section*{Author contributions statement} 
All authors contributed equally to this work.

\section*{Conflicts of interest} 
The authors declare no conflict of interest.

\section*{Data availability} 
No data was used in this investigation.

\end{document}